\documentclass[letterpaper,10pt]{iopart}

\usepackage{iopams,setstack,amsopn,amssymb} 

\usepackage{times}

\expandafter\let\csname equation*\endcsname\relax
\expandafter\let\csname endequation*\endcsname\relax
\usepackage{amsmath}

\usepackage{wrapfig}
\usepackage{etoolbox} 
\usepackage{graphicx}
\usepackage{cite}
\usepackage{color}
\usepackage{hyperref}
\usepackage[margin=1in]{geometry}
\usepackage{float}
\usepackage{listings}
\usepackage{caption}
\usepackage{xspace}
\usepackage[perpage, symbol]{footmisc}

\usepackage[scaled=0.9]{FiraMono}

\hypersetup{
    colorlinks=true,
    linkcolor=blue,
    citecolor=magenta,
    filecolor=magenta,      
    urlcolor=cyan
    }

\definecolor{codegreen}{rgb}{0,0.6,0}
\definecolor{codegray}{rgb}{0.5,0.5,0.5}
\definecolor{backcolor}{rgb}{0.98,0.98,0.98}

\definecolor{deepgreen}{rgb}{0.02,0.4,0.03}
\definecolor{deepblue}{rgb}{0.4,0.27,0.69}

\lstdefinestyle{codeblock}{
	backgroundcolor=\color{backcolor},   
	commentstyle=\color{codegreen},
	keywordstyle=\color{deepblue},
	numberstyle=\tiny\color{codegray},
	stringstyle=\color{deepgreen},
	basicstyle=\footnotesize,
	escapechar=\¢,escapebegin=\color{purple}, 
	otherkeywords={with},
	breakatwhitespace=false,         
	breaklines=true,      
	lineskip=1pt,
	captionpos=b,                    
	keepspaces=true,
	language=Python,
	numbers=none,                    
	numbersep=6pt,                  
	showspaces=false,                
	showstringspaces=false,
	showtabs=false,                  
	tabsize=2,
	frame=single,
	rulecolor=\color{lightgray},
	basicstyle=\fontencoding{T1}\ttfamily\footnotesize
}
\lstdefinestyle{output}{
	backgroundcolor=\color{white},   
	commentstyle=\color{codegreen},
	keywordstyle=\color{deepblue},
	numberstyle=\tiny\color{white},
	stringstyle=\color{deepgreen},
	basicstyle=\footnotesize,
	escapechar=\¢,escapebegin=\color{purple}, 
	otherkeywords={with},
	breakatwhitespace=false,         
	breaklines=true,      
	lineskip=1pt,
	captionpos=b,                    
	keepspaces=true,
	language=Python,
	numbers=none,                    
	numbersep=6pt,                  
	showspaces=false,                
	showstringspaces=false,
	showtabs=false,                  
	tabsize=2,
	frame=none,
	framerule=0pt,
	basicstyle=\fontencoding{T1}\ttfamily\footnotesize
}
\newcommand*{\ipythonpromptIn}[1]{\makebox[0pt][r]{\color{gray}\texttt{In:\;\;\;}\hspace{1em}}}
\newcommand*{\ipythonpromptOut}[1]{\makebox[0pt][r]{\color{gray}\texttt{Out:}\hspace{1em}}}

\graphicspath{{images/}}

\begin{document}
\title{Computer-aided quantization and numerical analysis of superconducting circuits}

\author{Sai Pavan Chitta$^1$, Tianpu Zhao$^1$, Ziwen Huang$^{1,2}$, Ian Mondragon-Shem$^1$, Jens Koch$^1$}
\address{$^1$ Department of Physics and Astronomy, Northwestern University, Evanston, Illinois 60208, USA}
\address{$^2$ \emph{current address:} Superconducting Quantum Materials and Systems Center,
Fermi National Accelerator Laboratory (FNAL), Batavia, IL 60510, USA}

\eads{saichitta2025@u.northwestern.edu}

\date{\today}

\begin{abstract}
The development of new superconducting circuits and the improvement of existing ones rely on the accurate modeling of spectral properties which  are key to achieving the needed advances in qubit performance. Systematic circuit analysis at the lumped-element level, starting from a circuit network and culminating in a Hamiltonian appropriately describing the quantum properties of the circuit, is a well-established procedure, yet  cumbersome to carry out manually for larger circuits. We present work utilizing symbolic computer algebra and numerical diagonalization routines versatile enough to tackle a variety of circuits. Results from this work are accessible through a newly released module of the \texttt{scqubits} package.
\end{abstract}

\section{\label{sec:level1}Introduction}

The past two decades have seen dramatic progress in the design and control of quantum devices based on superconducting circuits \cite{Gambetta2017, Wendin2017, Blais2021}. A central part of this effort is the design of qubits which are robust against decoherence and can be controlled and measured with high fidelity \cite{Brooks2013, Gyenis2021b, Zhang2021}. The systematic discovery of new qubits requires exploring the wealth of possible circuits formed by combining inductors, capacitors, and Josephson junctions as the basic circuit elements. This has led to the design of a growing number of qubit types, each with its own advantages  \cite{Nakamura1999,Orlando1999,Koch2007,Manucharyan2009,Brooks2013,Smith2020,Gyenis2021a}.  The task of exploring new circuit designs, however, can be tedious and become numerically challenging since the Hilbert space dimension needed to describe the low-energy physics of the system tends to grow exponentially with the number of elements in the circuit. In order to  make progress in the design of devices of increasing complexity, it is essential to promote tools that facilitate the efficient analysis of spectra and related key properties of superconducting circuits.

Several prescriptions have been proposed to approximate the low-energy Hamiltonian of a superconducting circuit, each depending on the level of detail needed for an accurate description \cite{Nigg2012,Bourassa2012,Solgun2014,Smith2016,Ding-Dawei,Minev2021}. For example, several approaches and software tools model the full geometry and distributed nature of the circuit \cite{Nigg2012,Solgun2014,Minev2021}. This is computationally costly, and often works most readily for circuits with only weak anharmonicity. The most basic, yet quite widely applicable quantization framework is based on the lumped-element description of a circuit \cite{Yurke1984,Devoret1995, Vool_2017,Burkard2004,Burkard2005,Ding-Dawei}. Despite the fact that this approach is the simplest one, it nonetheless presents challenges once the number of circuit elements is increased beyond the bare minimum of two (found, for example, in the transmon qubit). This has led to the development of software packages that aim to automate the construction of the Hamiltonian \cite{Charles2013,Parra2014,Gely2020,Aumann2021}.

In this work, we present a streamlined computer-aided approach to quantize lumped-element systems. By examining the circuit topology, we show how to classify and systematically detect four different types of degrees of freedom in a user-specified circuit: free, frozen, periodic and extended degrees of freedom. The identification of these variables can help reduce the number of degrees of freedom needed to describe the system and makes manifest the basis that should be used for an efficient representation of the Hamiltonian. This functionality has recently been incorporated into the \texttt{scqubits} Python package \cite{Groszkowski2021}, leveraging the full array of capabilities already available. We complement our approach with a hierarchical-diagonalization algorithm. This further mitigates the circuit-size limitations governed generally by the size of the Hamiltonian matrix which describes the low-energy spectrum of interest. 

For illustration, consider the task of finding the circuit Hamiltonian and numerical eigenenergies of a KITE circuit \cite{Smith2022}. Using a simple input file \verb|KITE.yaml| that encodes the circuit graph and associated circuit-element parameters (see Appendix \ref{sec:KITE_input}), generating the circuit Hamiltonian and performing numerical diagonalization is readily achieved by a few lines of code:
\begin{lstlisting}[language=Python, escapeinside={(*}{*)}, xleftmargin=35pt, xrightmargin=5pt]
(*\ipythonpromptIn{1}*)import scqubits as scq
kite = scq.Circuit.from_yaml("KITE.yaml", ext_basis="harmonic")
kite.sym_hamiltonian()
\end{lstlisting}
\begin{lstlisting}[language=Python, style=output, escapeinside={(*}{*)}, xleftmargin=35pt, xrightmargin=5pt, basicstyle=\small]
(*\ipythonpromptOut{1}{$ (15.7 Q_{1}^{2} + 26.4 Q_{1} Q_{2} - 52.8 Q_{1} Q_{3} + 26.4 Q_{2}^{2} - 52.8 Q_{2} Q_{3} + 52.8 Q_{3}^{2}) + (0.18 \Phi_{1}^{2} + 0.72 \Phi_{1} \theta_{1} - 0.36 \Phi_{1} \theta_{2} + 0.36 \Phi_{1} \theta_{3} + 0.18 \Phi_{2}^{2} 
+ 0.72 \Phi_{2} \theta_{1} - 0.36 \Phi_{2} \theta_{2} + 1.9 \theta_{1}^{2} - 1.44 \theta_{1} \theta_{2} + 1.18 \theta_{1} \theta_{3} + 0.36 \theta_{2}^{2} - 0.36 \theta_{2} \theta_{3} + 0.295 \theta_{3}^{2} - 5.9 \cos{\left(\theta_{2} \right)} - 5.9 \cos{\left(\theta_{2} + \theta_{3} \right)})$ *)
\end{lstlisting}
\vspace*{2mm}
Once suitable parameters for Hilbert space truncation are chosen, the eigenenergies are obtained simply by
\vspace*{2mm}
\begin{lstlisting}[language=Python, escapeinside={(*}{*)}, xleftmargin=35pt, xrightmargin=5pt]
(*\ipythonpromptIn{2}*)kite.eigenvals()
\end{lstlisting}
\vspace{-.15cm}
\begin{lstlisting}[language=Python, style=output, escapeinside={(*}{*)}, xleftmargin=35pt, xrightmargin=5pt]
(*\ipythonpromptOut{2}*)array([4.17710271, 6.47610652, 7.24813846, 8.47986653, 8.713839, 9.19210006])
\end{lstlisting}

The outline of the paper is as follows. In Section \ref{sec:generalized_coordinates} we set the stage for  the conceptual discussion of our approach by  reviewing the lumped-element description of superconducting circuits. This includes a simple discussion of constructing the circuit Lagrangian in terms of node variables.  In Section \ref{sec:degree_freedom}, we show how to systematically identify the four types of degrees of freedom in circuits consisting of capacitances, inductances,\footnote{Terminology employed here refers to the ideal circuit elements by ``capacitance" and ``inductances", whereas ``capacitor" and ``inductor" are understood as real building blocks that will include parasitic inductance and capacitance.} and Josephson junctions. In Section \ref{sec: algorithm}, we describe how the identification of these variables is implemented in the \texttt{scqubits} package to efficiently represent the Hamiltonian of a user-provided circuit. Finally, Section \ref{sec:conclusions} concludes with a summary and outlook of future developments. 

\section{\label{sec:generalized_coordinates} Circuit Lagrangian and classification of variables}

\subsection{Brief review of circuit Lagrangians}
The common starting point for the quantization of a superconducting circuit is the construction of a circuit Lagrangian \cite{Devoret1995,PhysRevB.69.064503, PhysRevA.29.1419, Girvin2011, Vool_2017, Rasmussen2021}. In this paper, we exclusively focus on  lumped-element circuits which are composed of capacitances, inductances, and Josephson tunnel junctions. (For more general discussions including additional elements such as gyrators, or extending the discussion beyond the lumped-element regime, consult Refs.\ \cite{Rymarz_2021,Egusquiza,Parra-Rodriguez, Riwar2022}.) Although seemingly restrictive, this small set of circuit elements spans a large family of different circuits, and includes the majority of superconducting qubits studied in the past, and used today. Each circuit element forms a branch in the connected circuit network.

Formulation of the Lagrangian depends on the definition of an appropriate set of \textit{generalized coordinates} (which we also refer to as \textit{variables}). In circuits including Josephson junctions, generalized-flux variables are the most convenient choice. For each branch $b,$ one defines
\begin{equation}
	\Phi_b(t) = \int_{t_0}^{t} d\tau\, V_b(\tau),\label{Eq_Phib}
\end{equation}
where $t_0$ is a reference time\footnote{In some references, $t_0$ is chosen as $-\infty$. To avoid questions about convergence of the integral, we choose $t_0$ to be finite and denote a time in the ``distant'' past when no external fields were present.} and $V_b(t)$ is the voltage across  branch $b$. The energy stored in a branch $b$ is given by the
\begin{align*}
{\text{\textit{charging}}\atop\text{\raisebox{3mm}{\textit{energy:}}}}\;\;  E =  \frac{C_b\dot{\Phi}^2_b}{2}, \qquad
    {\text{\textit{inductive}}\atop\text{\raisebox{3mm}{\textit{energy:}}}}\;\; E = \frac{\Phi^2_b}{2L_b},\qquad
    {\text{\textit{junction}}\atop\text{\raisebox{3mm}{\textit{energy:}}}}\;\;  E = -E_{Jb} \cos(2\pi \Phi_b/\Phi_0), 
\end{align*}
 for a capacitance $C_b$, inductance $L_b$, and junction with Josephson energy $E_{Jb}$, respectively.
($\Phi_0=h/2e$ is the flux quantum.)

The collection of all branch variables $\Phi_b(t)$ is in general not a set of independent variables: for each elementary loop $\mathsf{L}_\ell$ in the circuit, Kirchhoff's loop rule in conjunction with fluxoid quantization \cite{Tinkham2004, Vool_2017} imposes a constraint on the corresponding branch fluxes:
\begin{equation}
	\Phi_{\text{ext}}^\ell = \sum_{b\in \mathsf{L}_\ell}(\pm)_b\Phi_b(t). \label{Eq_SimpConstr}
\end{equation}
Here, $\Phi_\text{ext}^\ell$ is the external magnetic flux through the $\ell$-th loop, and the branch sign $(\pm)_b$ is determined according to the directionality of the branch relative to the chosen loop orientation.

\begin{figure}
    \centering
	\includegraphics[width=0.8\linewidth]{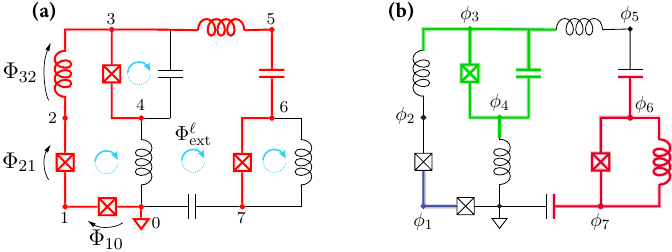}
	\caption{Example of a lumped-element superconducting circuit. (a) One choice of a spanning tree (marked in red) for the circuit, connecting the ground node with each other node by a unique path. (Only a subset of labels for branch variables and external fluxes are shown.)  (b) For the same circuit, coloring here indicates: a capacitively-coupled subcircuit or isolated island (blue), a junction-coupled subcircuit or superconducting island (red), and an inductively-coupled subcircuit (green). Labels denote the node variables.}
	\label{fig:example spanning tree}
\end{figure}

The construction of a spanning tree provides a systematic way to  determine a set of independent branch variables consistent with the constraints Eq.~(\ref{Eq_SimpConstr}). A spanning tree $\mathsf{T}$ is a subset of connected branches such that every circuit node is connected to ground by one, and only one, path within the spanning tree \cite{Vool_2017} [see the example in Fig.~\ref{fig:example spanning tree}(a)].\footnote{Spanning trees are not unique. This reflects the fact that the constraints \eqref{Eq_SimpConstr} leave open the choice of which branch variable in a loop to eliminate.} Each branch not part of the spanning tree is a closure branch. Iterative inclusion of closure branches, i.e., starting from the spanning tree, adding the first closure branch, then the second, etc., leads to the closing of one circuit loop at a time.\footnote{Careful analysis of the circuit in Fig.\ \ref{fig:example spanning tree} illustrates that there are cases in which the inclusion of the next closure branch appears to close multiple loops at a time. However, given the motivating purpose of resolving the loop constraints, it is always the newly closed loop containing exactly one closure branch which determines the relevant external loop flux. A more extensive discussion of this point has been presented in \cite{thomas2016analysis}} Once the constraints have been resolved in this way, the circuit Lagrangian can be expressed in terms of the branch variables $\Phi_{b\in\mathsf{T}}$ belonging to the spanning tree, with  $\Phi_{b\notin\mathsf{T}}$ eliminated by means of Eq.~\eqref{Eq_SimpConstr},
\begin{equation}\label{branchlagrangian}
    \mathcal{L} = T-V = \sum_{b}\left[\frac{1}{2}C_b\dot\Phi_b^2 - \frac{1}{2L_b}\Phi_b^2 + E_{J,b}\cos(2\pi\Phi_b/\Phi_0)\right]\bigg|_{b\notin \mathsf{T}:\text{ replace $\Phi_b$ via Eq.\ \eqref{Eq_SimpConstr}}}.
\end{equation}
Here, notation follows a simple scheme assigning the capacitance $C_b$ to branch $b=(n',n)$ if a capacitance indeed connects nodes $n$ and $n'$. Otherwise, the term is omitted from the summation. Analogous statements hold for inductance and junction contributions to the potential energy. We note in passing that time-dependent external flux renders the holonomic constraint \eqref{Eq_SimpConstr} among the coordinates time-dependent, and hence also yields a constraint among the generalized velocities $\dot\Phi_b$ \cite{You_Sauls_Koch_2019,Riwar2022}; for simplicity we limit ourselves to the time-independent case here.

Node variables present a useful alternative to branch variables. The node variable $\phi_n$ at node $n$ corresponds to the generalized flux obtained by adding up branch fluxes along the spanning tree; specifically, the spanning tree defines one path $\mathsf{P}_n$ connecting the ground node to node $n$. Then, the node variable $\phi_n$ is obtained by summing the branch variables along that path\footnote{Note: while the voltage across branch $b$ is well-defined, one cannot generally fall back to a description in terms of electric-potential differences. In the presence of non-zero external magnetic flux, even if constant in the present, the magnetic field had to be switched on after $t_0$ and be ramped up to its present value. As a result, the integral in equation \eqref{Eq_Phib} necessarily extends over a time interval during which the electric field was not curl-free, and could not be represented as a gradient field of the electric potential.}:
\begin{equation}
	\phi_n = \sum_{b \in \mathsf{P}_n} \Phi_{b},
\end{equation}
as illustrated in Fig.~\ref{fig:example spanning tree}(a). Branch variables within the spanning tree as well as the generalized flux associated with each closure branch can then be expressed in terms of node variables via 
\begin{equation}
		\Phi_{b\in \mathsf{T}} = \phi_{n} - \phi_{n'}, \qquad
		\Phi_{b\notin \mathsf{T}} = \phi_{n} - \phi_{n'} + \Phi_{\text{ext}}^\ell.
\end{equation}
Here $n, n'$ are the nodes connected by the branch $b=(n,n')$, and $\Phi_{\text{ext}}^\ell$ is the external flux threading the loop $\ell$ formed by the closure branch $b$ and the spanning tree. The circuit Lagrangian \eqref{Eq_SimpConstr} can then be re-expressed in terms of node variables in the general form
\begin{equation}
	\mathcal{L}(\dot{\boldsymbol{\phi}}, \boldsymbol{\phi}) = \frac{1}{2}\dot{\boldsymbol{\phi}}^\intercal \mathbf{C} \,\dot{\boldsymbol{\phi}} - V(\boldsymbol{\phi}),
\end{equation}
where $\boldsymbol{\phi}^\intercal=(\phi_1, \ldots, \phi_{N})$, and the $N\times N$ capacitance matrix $\mathbf{C}$ has 
non-zero off-diagonal entries $(\mathbf{C})_{nn'}=-C_b$ whenever a capacitance $C_b$ connects nodes $n\not=n'$;  diagonal entries are given by $ (\mathbf{C})_{nn}=-\sum_{m\neq n}(\textbf{C})_{n m}$. The potential energy in node variables is obtained via Eq.\ \eqref{branchlagrangian} by substitution, 
\begin{equation}
    V(\boldsymbol\phi) = \sum_{b=(n,n')}\left[\frac{1}{2L_b}(\phi_n'-\phi_n +\mathfrak{cl}_b\,\Phi_\text{ext}^{\ell_b})^2 - E_{J,b}\cos\left(\frac{2\pi}{\Phi_0}[\phi_n'-\phi_n+\mathfrak{cl}_b\,\Phi_\text{ext}^{\ell_b}]\right)\right],
\end{equation}
where $\mathfrak{cl}_b$ is the indicator for closure branches, i.e.\ $\mathfrak{cl}_b=1$ for $b\notin\mathsf{T}$ and $\mathfrak{cl}_b=0$ otherwise.

In principle, one might now proceed with either the branch- or node-variable Lagrangian, perform a Legendre transformation to obtain the Hamiltonian, and subsequently impose canonical quantization rules. However, it is generally worthwhile (especially where the ultimate aim involves numerical treatment)  to consider a further transformation of variables which will achieve two important goals: (1) we may reduce the number of relevant variables, for example by identifying conserved quantities, and (2) we cleanly distinguish variables associated with different boundary conditions upon quantization. A suitable variable transformation can thus help make the quantum description of the circuit both more transparent and numerically efficient.

\renewcommand{\arraystretch}{2.0} 
\begin{table}
    \footnotesize
    \sffamily
    \center
    	\includegraphics[width=1.0\linewidth]{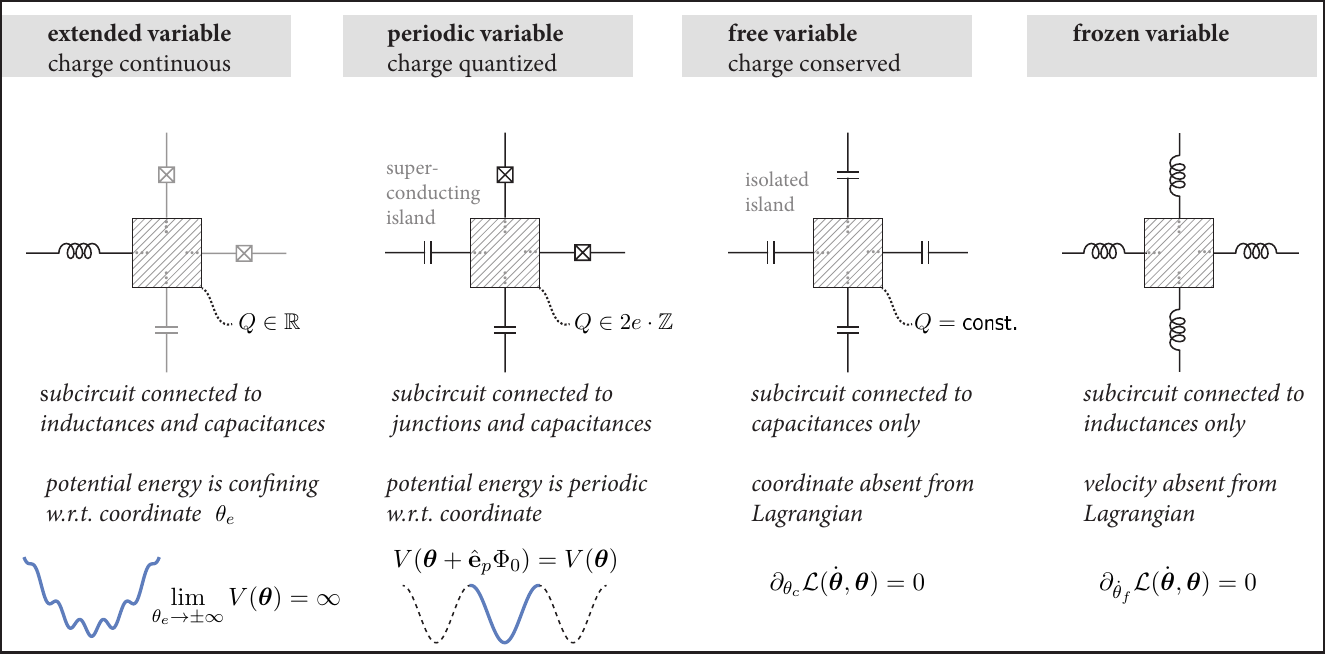}

	\caption{\label{table: variable type} Overview of the different types of variables. Classification is stated readily in the Lagrangian framework: a variable is \emph{extended} if the potential is confining, and \emph{periodic} if the potential is periodic along variable's axis. A variable (or coordinate) absent from the Lagrangian is free, while a generalized velocity absent leads to a \emph{frozen} variable. Each variable category is associated with specific conditions on how the corresponding subcircuit (hatched square) connects to the remaining circuit. The total charge $Q$ on the subcircuit is continuous for extended variables, quantized in units of $2e$ for periodic variables, and conserved in case of a free variable.}

\end{table}

\subsection{\label{sec:degree_freedom}Classification of variables}

Any variable $\theta$, no matter whether branch, node, or collective variable of neither type, can be classified based on four categories:
\begin{itemize}
    \item \textbf{Extended: } a variable $\theta_e$ is extended if it describes a direction in configuration space in which the potential energy is confining, expressed as $\lim_{\theta_e\rightarrow \pm \infty}V(\boldsymbol{\theta})=\infty$.
    \item \textbf{Periodic: } a variable $\theta_p$ is periodic if it aligns with a direction in configuration space in which the potential energy is periodic, i.e., $V(\boldsymbol{\theta}+\Phi_0 \mathbf{e}_p)=V(\boldsymbol{\theta})$. Here, $\mathbf{e}_p$ is the unit vector pointing along the axis associated with the $\theta_p$ coordinate. We exclude the case of a variable in which the potential energy is constant; this situation is covered by the following category.
    \item \textbf{Free:} a variable $\theta_c$ is free\footnote{Another common, though potentially misleading, name for this variable type is ``cyclic.''} if $\partial_{\theta_c} \mathcal{L}(\dot{\boldsymbol{\theta}},\boldsymbol{\theta})=0$. This means that $\theta_c$ does not appear in the potential energy, and the corresponding conjugate charge $Q_c=\partial_{\dot{\theta}_c} \mathcal{L}(\dot{\boldsymbol{\theta}},\boldsymbol{\theta})$ is conserved (the subscript $c$ refers to constant charge).
    \item \textbf{Frozen:} a variable $\theta_f$ is frozen if $\partial_{\dot{\theta}_f} \mathcal{L}(\dot{\boldsymbol{\theta}},\boldsymbol{\theta})=0$. In this case $\dot\theta_f$ is missing from the kinetic energy. This somewhat pathological case only arises when too many parasitic capacitances are neglected in the lumped element model. In the simplest setting, a frozen variable leads to an additional constraint given by $\partial_{\theta_f} \mathcal{L}(\dot{\boldsymbol{\theta}},\boldsymbol{\theta})=0$.
\end{itemize}

\begin{wrapfigure}{r}{0.27\textwidth}
\includegraphics[width=0.26\textwidth]{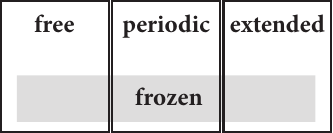}
\end{wrapfigure}The diagram on the right shows the relation among these four categories. The \emph{free}, \emph{periodic},  and \emph{extended} categories are mutually exclusive properties of the potential energy. The \emph{frozen} category regards the kinetic energy and as such can occur for each of the former three categories.

This classification holds several advantages when analyzing the Hamiltonian of a circuit. Recognizing free and frozen variables helps reduce the number of degrees of freedom needed to describe the system. Furthermore,  knowledge of periodic and extended variables guides a consistent and efficient choice of basis respecting the appropriate boundary conditions. 
Neither branch nor node variables are generally the suitable choice which recognizes the full number of periodic, free, and frozen degrees of freedom.  Failure to find all the variables of each type makes diagonalization of the quantum Hamiltonian usually more difficult, and can potentially lead to wrong conclusions about offset-charge dependence and charge-noise sensitivity of circuit eigenstates.

\section{Adaptive variable transformation}
In the following, we show how to systematically construct a linear variable transformation which will pinpoint all periodic, free, and frozen degrees of freedom. Against naive instinct, it is helpful to define this transformation by expressing the ``old" node variables in terms of the new variables, i.e., $\phi_n=\sum^{N}_{m=1}z_{nm}\theta_m$, where $\mathbf{Z}=(z_{nm})$ is an $N\times N$ invertible transformation matrix.  

\begin{figure}
\centering
    \includegraphics[width=0.75\textwidth]{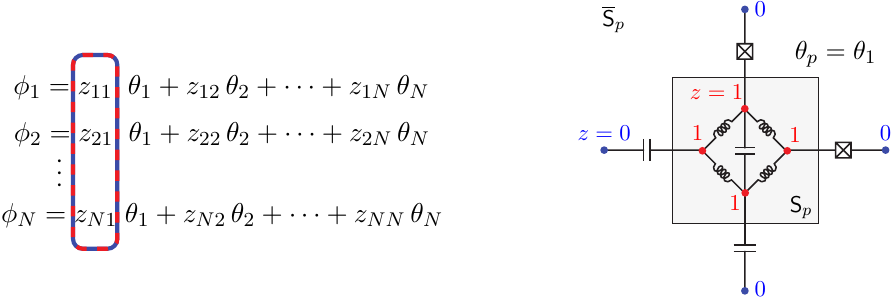}
    \caption{Variable transformation, and example of a junction-coupled subcircuit $\mathsf{S}_p$ and its complement $\overline{\mathsf{S}}_p$.\label{Fig_subcircuit}}
\end{figure}

We determine the $z_{nm}$ by studying the subcircuits of the system, defined as connected sub-graphs of the circuit. We distinguish three relevant classes of subcircuits: (1) A subcircuit forms an \textit{isolated island} if it is connected to the rest of the circuit exclusively by capacitances; (2)  a subcircuit forms a \textit{superconducting island} if it is coupled to the remainder of the circuit by capacitances and at least one junction (but not via any inductances); (3) a subcircuit is purely inductively coupled if it connects to the rest of the circuit only through inductances. Examples of each type of subcircuit are shown in Fig.~\ref{fig:example spanning tree}(b).

\subsection{Periodic and extended variables}

For the analysis of periodic variables, consider a subcircuit $\mathsf{S}_p$ forming a \textit{superconducting island} as in the example of Fig.\ \ref{Fig_subcircuit}. There, $\mathsf{S}_p$ is shaded in gray and couples to its circuit complement $\overline{\mathsf{S}}_p$ exclusively through capacitances and junctions.  Any inductances present in the circuit are entirely inside  $\mathsf{S}_p$ or inside $\overline{\mathsf{S}}_p$, but not at their interface. Given this partitioning, we see there is a simple choice of coefficients $z_{n1}$ that will render the variable $\theta_1$  periodic, namely:
\begin{equation}
    z_{n1} = 1\text{ for all }n\in\mathsf{S}_p,\quad z_{n1} = 0\text{ for all }n'\in\overline{\mathsf{S}}_p. 
\end{equation}
In other words, a change in the new flux variable $\theta_1$ will simultaneously raise the flux of each node variable inside the superconducting island by equal amounts, but keep the flux on all nodes outside the superconducting island unchanged. Since $\theta_1$ clearly produces no flux difference across any inductance, the potential energy
must indeed be periodic with respect to $\theta_1$.

The full set of periodic variables is obtained by identifying all superconducting islands in the circuit. This indicates that the full number of periodic variables, $N_p$, is equal to the number of \textit{irreducible} superconducting islands $\mathsf{S}_p$ which cannot be decomposed into smaller superconducting islands. Once all periodic variables are assigned, any remaining variable $\theta_e$ must now appear in the quadratic potential energy terms generated by the inductances. These terms have the usual confining property  $\lim_{\theta_e\rightarrow \pm \infty}V(\boldsymbol{\theta})=\infty$. As a result, all these degrees of freedom are extended.

\subsection{\label{sec:degree_freedom:free_periodic}Free variables } 

Free variables are treated in a manner nearly identical to the periodic case. Suppose $\theta_c=\theta_{1}$ is one of the free variables. In this case, changes in $\theta_c$ cannot produce node-flux differences across any of the inductances or junctions. Again, partitioning into subcircuits facilitates the definition of new variables. For free variables, the subcircuit $\mathsf{S}_c$ must now be an \emph{isolated island} connected to the remainder of the circuit $\overline{\mathsf{S}}_c$ via capacitances exclusively. Carrying over the definition of $z_{n1}$ to the isolated island case,
\begin{equation}
    z_{n1} = 1\text{ for all }n\in\mathsf{S}_c,\quad z_{n1} = 0\text{ for all }n\in\overline{\mathsf{S}}_c, 
\end{equation}
equips us with a free variable: since flux differences generated by $\theta_1$ are across capacitances only, $\theta_1$ will be absent from all potential energy terms associated with inductances and junctions. As a result, $\partial_{\theta_c}V(\boldsymbol{\theta})=0$, so $\theta_c$ is indeed free. The full number of free variables, $N_c$, corresponds to the total number of irreducible isolated islands in the given circuit.

\subsection{\label{sec:degree_freedom:frozen_degree}Frozen variables}

As the last category, we discuss the case of frozen variables. For a frozen variable, the relevant subcircuit is of type (3): a subcircuit coupled to the remaining circuit through inductances only. Once such a subcircuit is identified, a frozen variable $\theta_f$ is readily defined following the now familiar way: $z$-coefficients for nodes inside and outside the subcircuit are set to 1 and 0, respectively. As a result, the only flux differences that occur under a change in $\theta_f$ are across inductance terms in the Lagrangian. Kinetic energy terms due to capacitances remain unchanged such that $\partial_{\dot{\theta}_f}\mathcal{L}=0$, and $\theta_f$ is, hence, indeed a frozen variable. The full number of frozen variables corresponds to the total number of irreducible, purely inductively coupled subcircuits.

A few remarks about frozen variables are in order. Variables only appear frozen in situations where too many parasitic capacitances have been dropped from the description and the capacitance matrix has become singular. A foolproof, physically accurate but not necessarily efficient way of handling this situation is to re-admit the omitted parasitic capacitances which physically are always present. This is not always desirable as it generally increases the number of degrees of freedom retained in the modeling of the circuit. 

Dropping a parasitic capacitance from the charging-energy contributions is akin to setting this capacitance to zero. In the mechanically analogous system, this corresponds to setting the mass of a particle to zero. The $C\to0$ limit is generally a singular limit \cite{Bender1978} that exhibits subtleties particularly significant when junction capacitances are omitted \cite{Rymarz_2021}. We exclude this case here and focus on the more benign scenario in which one of the Euler-Lagrange equations turns into a linear holonomic constraint for $C=0$, see Subsection \ref{sec:eliminaton}.

\subsection{Full transformation}

As an example, consider the circuit in Fig.~\ref{fig:example spanning tree}(b), which has three subcircuits. The corresponding transformation is 
\begin{equation}
	\renewcommand\arraystretch{1}
	\begin{pmatrix}
		\phi_1 \\
		\phi_2 \\
		\phi_3 \\
		\phi_4 \\
		\phi_5 \\
		\phi_6 \\
		\phi_7 \\
	\end{pmatrix}
	= \begin{pmatrix}
	0 & 0 & 1 &  \\[-0.7em]
	0 & 0 & 0 & \vdots \\
	0 & 1 & 0 & \\
	0 & 1 & 0 & \,\,\cdots \,\,\, z_{nm}\,\,\,\cdots \,\, \\
	0 & 0 & 0 & \\[-0.7em]
	1 & 0 & 0 & \vdots\\
	1 & 0 & 0 &  
	\end{pmatrix}   
	\begin{pmatrix}
		\theta_{c} \\
		\theta_{\!f} \\
		\theta_{p} \\
		\theta_{e,1}   \\
		\theta_{e,2}   \\
		\theta_{e,3}   \\
		\theta_{e,4}   
	\end{pmatrix}.
	\label{eqn: example transformation}
\end{equation}
This transformation identifies a free, a frozen and a periodic variable. The remaining variables are extended. The matrix elements corresponding to the extended variables $\{\theta_{e,i}\}_{i=1}^4$ are chosen such that the transformation is invertible.  The Lagrangian in terms of the new variables is
\begin{equation}
    \mathcal{L}(\dot{\boldsymbol{\theta}},\boldsymbol{\theta}) = \tfrac{1}{2}\dot{\boldsymbol{\theta}}^\intercal \widetilde{\mathbf{C}}\,\dot{\boldsymbol{\theta}} - V(\boldsymbol{\theta}),\label{Eq_LagrZ}
\end{equation}
where  $\widetilde{\mathbf{C}}=\mathbf{Z}^\intercal \mathbf{C}\, \mathbf{Z}.$

\subsection{Elimination of non-dynamical variables\label{sec:eliminaton}}

To obtain the circuit Hamiltonian, we first eliminate the frozen and free variables. Eliminating frozen variables is, in fact, a necessary step, since the capacitance matrix [Eq.~(\ref{Eq_LagrZ})] is singular in the presence of frozen variables.  

To carry out the elimination of frozen variables from the Lagrangian, we use that $\partial_{\theta_{\!f,i}} \mathcal{L}=\frac{d}{dt}\partial_{\dot{\theta}_{\!f,i}} \mathcal{L}=0,$ where $i=1,\ldots, N_f$ labels the inductively-coupled subcircuits. In general, this set of $N_f$ equations are nonlinear due to the cosine terms in the Lagrangian. As a crucial simplification, we assume that none of the junction capacitances are omitted in the description. In that case, only subcircuits with purely linear-inductive coupling to the remainder of the circuit lead to frozen variables. Thus, the variables $\{\theta_{\!f,i}\}$ can be solved for using the constraints emerging from the Euler-Lagrange equations. Upon eliminating the $\{\theta_{\!f,i}\}$ from the Lagrangian, the kinetic energy is expressed in terms of the invertible capacitance matrix $\widetilde{\mathbf{C}}_{cpe}$ obtained by removing the rows and columns in $\widetilde{\mathbf{C}}$ that were associated with frozen variables. At this point, we perform a Legendre transform to obtain the classical Hamiltonian
\[
H=\boldsymbol{Q}^\intercal_{cpe}\cdot\dot{\boldsymbol{\theta}}_{cpe}-\mathcal{L}(\dot{\boldsymbol{\theta}}_{cpe},\boldsymbol{\theta}_{pe}),\]
where $\boldsymbol{Q}_{cpe}=\partial_{\dot{\boldsymbol{\theta}}_{cpe}}\mathcal{L}$ and  $\boldsymbol{\theta}_{cpe}$ corresponds to free, periodic and extended variables.  
 
For every free variable $\theta_c$, the conjugate charge $Q_c$ is a conserved quantity. As shown in \ref{sec: App_routhian}, these conserved charges have no impact on the dynamics of the remaining degrees of freedom, and they may hence be omitted. The kinetic energy of the Hamiltonian can thus be written in the form $T=\frac{1}{2}\mathbf{Q}^\intercal_{pe}\widetilde{\mathbf{C}}^{-1}_{pe}\mathbf{Q}_{pe},$ where $\widetilde{\mathbf{C}}^{-1}_{pe}$ is the matrix block of $\widetilde{\mathbf{C}}^{-1}_{cpe}$ that corresponds to the extended and periodic variables only. Similarly, the vector $\mathbf{Q}_{pe}$ only includes the non-conserved charges conjugate to the variables $\boldsymbol{\theta}_{pe}.$

\subsection{Quantization}
Finally, canonical quantization yields the quantum mechanical Hamiltonian. Quantization is achieved by promoting  the coordinate and conjugate momenta to operators $\hat{\theta}_n$ and $\hat{Q}_n=-i\hbar\partial_{\theta_n}$ satisfying canonical commutation relations. The Hamiltonian finally takes the form
\begin{equation}
    \hat{H}=\tfrac{1}{2}(\hat{\mathbf{Q}}_{pe}-\mathbf{Q}_g)^\intercal \widetilde{\mathbf{C}}_{pe}^{-1}(\hat{\mathbf{Q}}_{pe}-\mathbf{Q}_g)+V(\hat{\boldsymbol{\theta}}_{pe}),
\end{equation}
where we have introduced a vector $\mathbf{Q}_g$ representing off-set charges. Since  periodic and extended degrees of freedom are identified explicitly, the boundary conditions required for the wavefunction $\Psi(\boldsymbol{\theta_{pe})}$ can be stated transparently:
\begin{align*}
{\text{\textit{Periodic directions:}}}\;\;  \Psi(\boldsymbol{\theta}_{pe}+\mathbf{e}_p\Phi_0)=\Psi(\boldsymbol{\theta}_{pe}), \qquad
    {\text{\textit{extended directions:}}}\;\; \lim_{\theta_e\rightarrow \pm\infty}\Psi(\boldsymbol{\theta}_{pe})=0.
\end{align*}
The basis choice is further simplified by the distinction between periodic and extended variables. Periodic degrees of freedom can be represented in the discrete charge basis, extended variables  in the phase basis or an appropriately chosen oscillator basis. The process of constructing the Hamiltonian $\hat{H}$ of a given circuit and obtaining its spectrum has been implemented in the \texttt{scqubits} package \cite{Groszkowski2021}, as we explain in the following.

\section{Implementation in \texttt{scqubits}\label{sec: algorithm}}

\begin{figure}
\centering
    \includegraphics[width=0.8\textwidth]{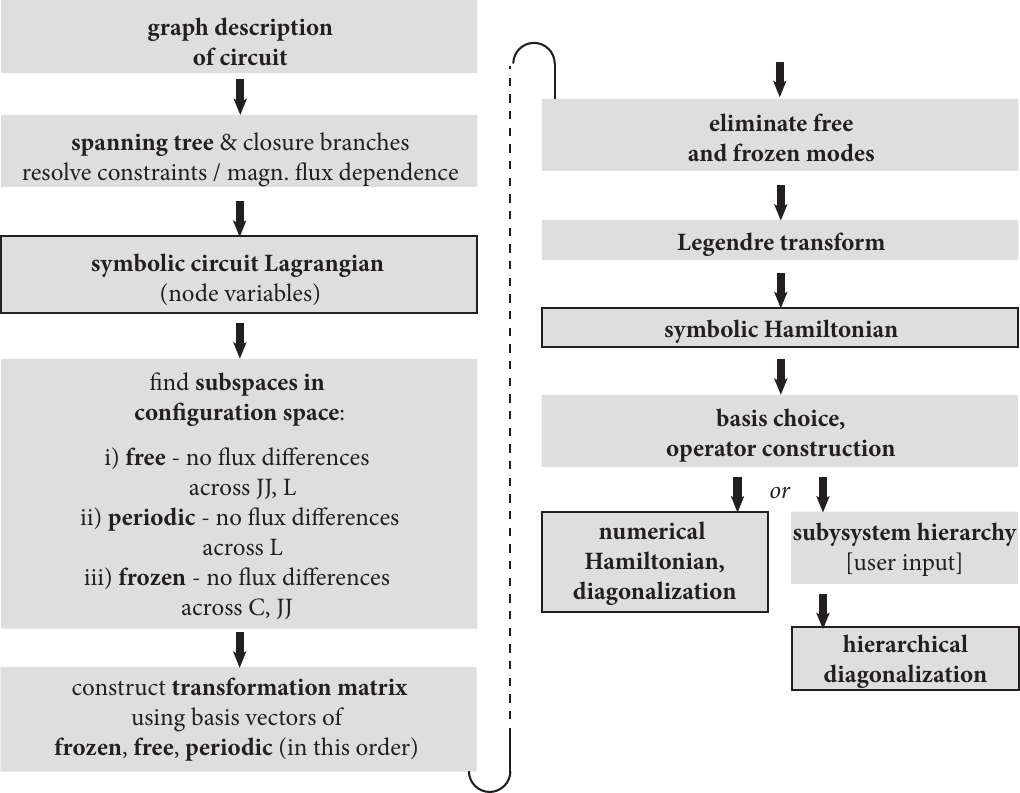}
    \caption{Flowchart outlining the implementation of computer-aided quantization and numerical diagonalization in the \texttt{scqubits} package.\label{flowchart}}
\end{figure}

\subsection{Variable transformation}
The variable transform discussed in Section \ref{sec:degree_freedom} is implemented programmatically in order to automate the quantization of custom superconducting circuits. Several steps described in terms of subcircuits are equivalently formulated in more compact linear-algebra language applied to the vector space of node variables. 

The algorithm proceeds by dividing the vector space of node variables into independent subspaces associated with the different variable categories.  For instance, the subspace corresponding to periodic variables is spanned by those vectors $\mathbf{z}_p=(z_{1p},z_{2p},\ldots,z_{Np})$ which are compatible with the condition of vanishing flux difference across all inductances. Similar  subspaces are constructed for free and frozen degrees of freedom present in the circuit. The full transformation matrix\footnote{Note that the transformation matrix generated can have any real valued coefficients. The only condition is that $z_{ip}$ - $z_{jp} \in \mathbb{Z}$ such that $i\in\mathsf{S}_p$ and $j \notin \mathsf{S}_p$, which ensures that the associated conjugate charge is periodic in a single Cooper pair. In some instances, having non-integer elements in the transformation matrix can help to increase the sparsity of the resultant matrix representation of the Hamiltonian, which can significantly reduce the diagonalization time\cite{Ding-Dawei}.} is obtained by combining the  vectors spanning the above three subspaces, and adding additional linearly independent vectors, if needed, to complete the basis. 

Fig.\ \ref{flowchart} summarizes the steps involved in this  algorithm which has been implemented as a new module extending the \texttt{scqubits} package \cite{Groszkowski2021}. The \texttt{Circuit} class added for this purpose facilitates the generation and manipulation of symbolic circuit Hamiltonians by leveraging the Python package \texttt{SymPy}.

\subsection{Numerical diagonalization}
To obtain the spectrum of the circuit numerically, its Hamiltonian must be represented as a matrix with respect to a suitable basis. We choose bases $\{| \mu_{i}\rangle\}$ ($i=1,2\ldots$) for all individual degrees of freedom. For the periodic degrees of freedom we employ the discrete charge eigenstates as a convenient (though not necessarily ideal) basis. For extended degrees of freedom, we either apply simple discretization with multi-point stencils turning derivatives into finite differences, or employ the harmonic oscillator basis associated with the potential energy obtained when omitting junction terms.

In the ordinary diagonalization scheme, the basis of the full Hilbert space is chosen as the  Kronecker product $|\mu_1\rangle\otimes|\mu_2\rangle \otimes \ldots$ For each degree of freedom, individual truncation levels must be monitored to ensure convergence of the procedure. By contrast, hierarchical diagonalization aims to mitigate the fact that the dimension of the Hilbert space grows exponentially with the number of nodes\footnote{Details of convergence depend not only on the node number, but also critically on the specific parameter regime determined by the circuit parameters.}. The \texttt{Circuit} class offers an implementation of the hierarchical diagonalization procedure following ideas inspired by Kerman \cite{Kerman_2020}. In the first stage,  subsystems are diagonalized individually to obtain low-lying energy eigenstates $\{\vert E_i \rangle\}.$ The basis of the full Hilbert space used in this case is the Kronecker product $|E_1\rangle\otimes|E_2\rangle \otimes \ldots$ Under favorable conditions of moderate hybridization among subsystems, this procedure lowers the overall Hilbert space dimension.  Once either of the two diagonalization schemes has been chosen, the full functionality of \texttt{scqubits} for visualizing spectra and wavefunctions, performing parameter sweeps, etc.\ is available to explore the behavior of the custom circuit.

As a concrete illustration of the hierarchical-diagonalization scheme, consider the fluxonium coupler circuit (see Appendix \ref{sec: fluxonium coupler}), composed of five nodes and four degrees of freedom. Simulation of the full circuit Hamiltonian generally calls for diagonalization of matrices of size $ 10^6\times 10^6$. In suitable parameter regimes of practical interest, the system can be meaningfully divided into weakly coupled sub-systems, and hierarchical diagonalization proves to be a key step that decreases  matrix sizes to $10^3\times 10^3$ and dramatically reduces the required runtime by two orders of magnitude.

\section{\label{sec:conclusions} Conclusions}

We described a computer-aided framework to quantize general lumped-element superconducting circuit systems. We showed how to systematically identify periodic, extended, and non-dynamical (free and frozen) degrees of freedom in a system. Eliminating the non-dynamical degrees of freedom leads to an efficient representation of the Hamiltonian. This framework is implemented programmatically in the new module \texttt{circuit} of \texttt{scqubits} Python library, which can simulate a user-defined circuit described as a graph. Hierarchical diagonalization was also implemented, which made numerical analysis for larger circuits accessible.

Future work will further extend the scope of the module by incorporating coherence-time calculations for custom circuits, handle the quantization and numerical analysis of circuits in the presence of time-dependent magnetic flux, and utilize the computation of perturbative corrections for hierarchical diagonalization discussed in Ref.\ \cite{Kerman_2020}. 
Additional opportunities for enhancing the performance of our code lie in optimizing the choice of variable transformation which is generally not completely fixed by the protocol reported here. The remaining freedom may offer routes to increasing numerical efficiency, especially if variable transformations can reduce explicit coupling terms \cite{Ding-Dawei}.

\section*{Note added}
During preparation of this manuscript, we were made aware of ongoing work in Amir Safavi-Naeini's group with a similar scope. Our two groups did not discuss details of our projects, but agreed to post preprints on the arXiv simultaneously.

\ack
{\small
We thank Andrew Houck, Zlatko Minev, Michel Devoret, Xinyuan You, Daniel Weiss and Venkat Chandrasekhar for valuable discussions. 
This material is based upon work in part supported by the U.S.\ Department of Energy, Office of Science, National Quantum Information Science Research Centers, Co-design Center for Quantum Advantage (C2QA) under contract number DE-SC0012704, and in part
 by the Army Research Office under grant W911NF-1910016. For this work, IMS was funded by the DOE, ZH and SPC were funded by the ARO, and JK was funded by ARO and DOE. During the writing and editing of the manuscript, ZH was funded by the U.S.\ Department of Energy, Office of Science, National Quantum Information Science Research Centers, Superconducting Quantum Materials and Systems Center (SQMS) under contract number DE-AC02-07CH11359. We thank the Air Force Office of Scientific Research for supporting the \texttt{scqubits} package under grant FA9550-20-1-0271.}

\appendix

\section{Examples}
\subsection{User-defined transformation for the KITE circuit\label{sec:KITE_input}}
The following string encodes the circuit graph for the example KITE circuit in Section \ref{sec:level1}. This string can be passed    to the method \texttt{from\_yaml} directly or through a text file. 

\begin{figure}[H]
    \includegraphics[width=0.75\textwidth]{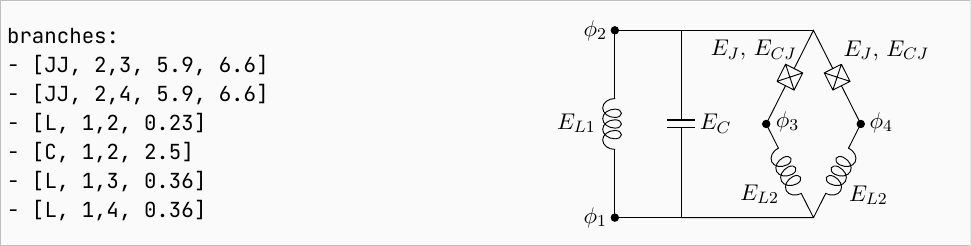}
\end{figure}

The parameters are (in units of GHz): $E_C = 2.5$, $E_{L1} = 0.23$, $E_{L2} = 0.36$, $E_{J} = 5.9$, $E_{CJ} = 6.6$. The following snippet sets the truncation for the Hilbert space, chooses a more efficient transformation following Ref.\ \cite{Smith2022} and also chooses the closure branches.

\begin{lstlisting}[language=Python, escapeinside={(*}{*)}, xleftmargin=35pt, xrightmargin=5pt, mathescape=true]
(*\ipythonpromptIn{1}*)trans_mat = np.array([[ -1,  -1,  -1,  1],
                       [ -1,  -1,  3,  1],
                       [ -1,  3, -1,  1],
                       [ 3,  -1,  -1,  1]])*0.25
closure_branches = [kite.branches[-2], kite.branches[-1]]
kite.configure(transformation_matrix=trans_mat, closure_branches=closure_branches)
kite.cutoff_ext_1 = 30
kite.cutoff_ext_2 = 30
kite.cutoff_ext_3 = 30
\end{lstlisting}

\subsection{Hierarchical diagonalization for two fluxonia with a tunable coupler \label{sec: fluxonium coupler}}

The following string for the input file \texttt{fluxonium-coupler.yaml} encodes the circuit graph for the fluxonium coupler \cite{DannyAPS, HelinAPS}:
\begin{figure}[H]
\includegraphics[width=1\textwidth]{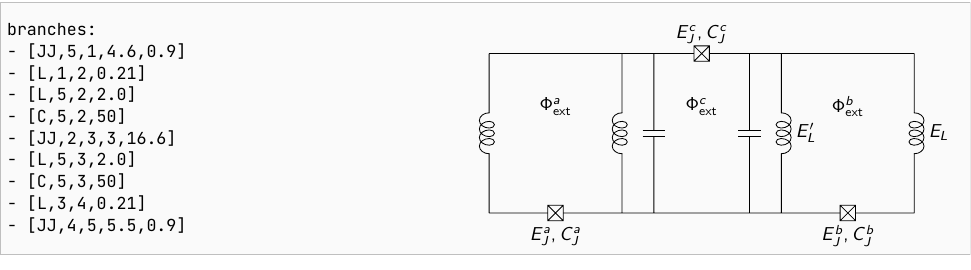}
\end{figure}

\noindent The circuit can be diagonalized using hierarchical diagonalization by the following lines of code:

\begin{lstlisting}[language=Python, escapeinside={(*}{*)}, xleftmargin=35pt, xrightmargin=5pt, mathescape=true, escapebegin=\color{black}]
(*\ipythonpromptIn{1}*)import scqubits as scq

fluxonium_coupler = scq.Circuit.from_yaml("fluxonium-coupler.yaml", ext_basis="harmonic", initiate_sym_calc=False, basis_completion="canonical")

system_hierarchy = [[1], [2], [3], [4]]
closure_branches = [fluxonium_coupler.branches[i] for i in [0, 4, -1]]

fluxonium_coupler.configure(closure_branches=closure_branches, system_hierarchy=system_hierarchy, 
                     subsystem_trunc_dims = [6, 6, 6, 6])
                     
fluxonium_coupler.(*$\Phi$*)1 = 0.5 + 0.01768
fluxonium_coupler.(*$\Phi$*)2 = -0.2662
fluxonium_coupler.(*$\Phi$*)3 = -0.5 + 0.01768

fluxonium_coupler.cutoff_ext_1 = 110
fluxonium_coupler.cutoff_ext_2 = 110
fluxonium_coupler.cutoff_ext_3 = 110
fluxonium_coupler.cutoff_ext_4 = 110

eigs = fluxonium_coupler.eigenvals()
eigs - eigs[0]
\end{lstlisting}
\begin{lstlisting}[language=Python, style=output, escapeinside={(*}{*)}, xleftmargin=35pt, xrightmargin=5pt, mathescape=true]
(*\ipythonpromptOut{1}*) array([0.        , 0.03559404, 0.05819727, 0.09378676, 4.39927874, 4.43488613]) 
\end{lstlisting}
\vspace*{2mm}
We have verified that these eigenvalues agree with results obtained independently by the authors of \cite{DannyAPS}.

\section{Role of conserved charges in the dynamics of superconducting circuits\label{sec: App_routhian}}

In this appendix, we show that conserved charges do not impact the physics of the dynamical variables.  To simplify the analysis, we will assume that there is a single  variable $\theta_1$ that is free. We perform a partial Legendre transform with respect to the free variable to obtain the Routhian function
\begin{equation}
\mathcal{R}(\dot{\boldsymbol{\theta}}_{pe}, \boldsymbol{\theta}_{pe}; Q_1)={Q}_{1}\dot{{\theta}}_{1}-\mathcal{L}(\dot{\boldsymbol{\theta}}, {\boldsymbol{\theta}}),   
\end{equation}
where ${Q}_{1}=\partial_{\dot{{\theta}}_{pe}}\mathcal{L}(\dot{\boldsymbol{\theta}},\boldsymbol{\theta})$ and $\boldsymbol{\theta}_{pe}$  represents the variables excluding $\theta_1.$ The Routhian can be written in the form
\begin{equation}
	\mathcal{R}(\dot{\boldsymbol{\theta}}_{pe},\boldsymbol{\theta}_{pe}; Q_1) = 
	-\mathcal{L}_0(\dot{\boldsymbol{\theta}}_{pe},\boldsymbol{\theta}_{pe})
	+ \frac{1}{2(\widetilde{\mathbf C}_{cpe})_{1, 1}}\left(\sum_{i\neq 1} (\widetilde{\mathbf C}_{cpe})_{i, 1} (\dot{ \boldsymbol{\theta}}_{pe})_{i}\right)^2
	+ \frac{d}{dt}G(Q_1,t),
\end{equation}
where $\mathcal{L}_0(\dot{\boldsymbol{\theta}}_{pe},\boldsymbol{\theta}_{pe})$ is the Lagrangian of the system without the free degree of freedom, and 
\begin{equation}
	G(Q_1,t)=-\frac{ Q_{1}}{(\widetilde{\mathbf C}_{cpe})_{1, 1}}\left(\sum_{i\neq 1} (\widetilde{\mathbf C}_{cpe})_{i, 1} ( \boldsymbol{\theta}_{pe})_{i}\right)
	+ \frac{Q_1^2}{2(\widetilde{\mathbf C}_{cpe})_{1, 1}}t.
\end{equation}
The dynamics of the variables $\boldsymbol{\theta}_{pe}$ is determined by the Euler-Lagrange equations $\frac{d}{dt}\partial_{\dot{\boldsymbol{\theta}}_{pe}}\mathcal{R}=\partial_{\boldsymbol{\theta}_{pe}}\mathcal{R}.$ Since the conserved charge $Q_1$ enters through $G(Q_1,t)$ as a total time derivative in the Routhian, the value of $Q_1$ has no impact on the dynamics. As a result, after obtaining the classical Hamiltonian, the charge $Q_1$ can be set to zero. The generalization to more than one free variable can be carried out straightforwardly.

\section*{References}

\providecommand{\newblock}{}

\end{document}